\documentclass[%
aip,rsi,amsmath,amssymb,
 reprint,]{revtex4-1}

\usepackage{graphicx}
\usepackage{dcolumn}
\usepackage{bm}
\usepackage{color}

\usepackage{titlesec} 
\titlespacing\section{0pt}{10pt}{4pt}
\titlespacing\subsection{0pt}{10pt}{2pt}

\begin{document}
\title{A surface plasmon platform for angle-resolved chiral sensing}
\author{Sotiris Droulias}\email{sdroulias@iesl.forth.gr}
\affiliation{Institute of Electronic Structure and Laser, FORTH, 71110 Heraklion, Crete, Greece}
\author{Lykourgos Bougas}\email{lybougas@uni-mainz.de}
\affiliation{Institut f\"ur Physik, Johannes Gutenberg Universit\"at-Mainz, 55128 Mainz, Germany}



\date{\today}

\begin{abstract}
Chiral sensitive techniques have been used to probe the fundamental symmetries of the universe, study biomolecular structures, and even develop safe drugs. As chiral signals are inherently weak and often suppressed by large backgrounds, different techniques have been proposed to overcome the limitations of traditionally used chiral polarimetry. Here, we propose an angle-resolved chiral surface plasmon resonance (CHISPR) scheme that can detect the absolute chirality (handedness and magnitude) of a chiral sample and is sensitive to both the real and imaginary part of a chiral sample's refractive index. We present analytical results and numerical simulations of CHISPR measurements, predicting signals in the mdeg range for chiral samples of $<$100\,nm thickness at visible wavelengths. Moreover, we present a theoretical analysis that clarifies how our far-field measurements elucidate the underlying physics. This CHISPR protocol does not require elaborate fabrication and has the advantage of being directly implementable on existing surface plasmon resonance instrumentation.
\end{abstract}

\maketitle

\section{Introduction}
Chirality is a fundamental property of life with far-reaching implications among all disciplines of science. Chiral sensitive techniques have enabled the study of fundamental symmetries of the universe\,\cite{Fortson1984}, determination of biomolecular structures\,\cite{Fasman2010,KELLY2005,Norder2010}, and development of safe and effective drugs\,\cite{Hutt1996,Nguyen2006}, to name few of the most prominent applications of chiral sensing. \\
\indent The polarimetric techniques of optical rotatory dispersion (ORD) and circular dichroism (CD) are among the most widely used research tools in modern science for chiral sensing\,\cite{barron_2004}. However, polarimetric measurements are typically challenging as the measured signals are small and often suppressed by large backgrounds. To overcome the limitations of traditional polarimetry in chiral sensing, different techniques have been proposed in the recent years. These techniques, which aim to enhance the chiral wave-matter interaction, can in principle be arranged into two main categories, as they rely mainly on either (a) path-length enhancement or (b) chiral-field enhancement. The first type of techniques is primarily cavity-based, for which the ORD and CD signals are enhanced by the number of cavity-passes (typically about 10$^3$-10$^4$)\,\cite{Poirson1998,Mueller2000,Sofikitis2014,Bougas2015}. However, cavity-enhanced techniques become inadequate in systems with losses originating from absorption and/or scattering (e.g. chiral molecules within complex matrices, thin films, liquid and/or solid systems), because losses hinder the path-length enhancement. The chiral-field-enhancement techniques rely primarily on generating probing-electromagnetic-fields with chiral densities higher than circularly polarized plane waves\,\cite{Hendry2010,Tang2011,Davis2013,Karimullah2015,Luo2017,Zhao2017,Mohammadi2018}. Chiral and achiral nanophotonic systems, such as plasmonic/dielectric nanostructures and metamaterials, can generate contorted intense near-fields with high chiral densities around a resonance frequency of the nanosystem, thus amplifying the chiral-chiral interactions between them and a molecule, resulting in up to $10^6$ enhanced response to weak molecular CD effects\,\cite{Hendry2010}. However, in most demonstrations, elaborate fabrication is required and the employed nanosystems have their own intrinsic chiroptical response that contributes in the total CD signal and, thus, inhibits the absolute and quantitative measurement of chirality\,\cite{Hendry2010,Tang2011,Davis2013,Karimullah2015,Luo2017,Zhao2017}.\\
\indent Considering the importance of chiral sensing, it is vital to develop alternative schemes that overcome the above-mentioned limitations of path-length enhancement and/or chiral-field enhancement techniques. In this article we show that surface plasmon resonance (SPR) allows for an absolute measurement of chirality (handedness and magnitude) of a chiral system. SPR has become an important technology in the areas of biochemistry, biology, and medical sciences because of its real-time, label-free, and noninvasive nature (see Refs.\,\cite{Nguyen2015,Schasfoort2017} and references therein). We demonstrate how chiral-sensitive SPR (CHISPR) is able to quantitatively detect both the real and imaginary part of the refractive index of a chiral substance (responsible for the refraction and absorption, respectively), contrary to most demonstrations that employ metallic nanostructures and/or metasurfaces. We show that CHISPR is particularly suitable for chiral sensing from thin samples which are not easily measurable using alternative techniques, and that our technique has the advantage of being applicable directly on existing SPR instrumentation without the need for additional elaborate fabrication.\\
\indent In particular, using analytical calculations and numerical simulations we show how the presence of a chiral substance on top of a metal results in angular shifts in SPR measurements, with which we are able to identify the sign and quantify the magnitude of the sample's chirality. Furthermore, we demonstrate how an appropriate polarimetric analysis of the outgoing reflected beam enables the absolute measurement of signals from thin chiral layers with mdeg signals at visible wavelengths. Finally, we discuss our results using a theoretical description based on the concept of optical chirality flux\,\cite{Lipkin1964,Tang2010,Bliokh2011,Coles2012,Philbin2013,Poulikakos2016} elucidating the underlying relation between the far-field detection scheme we employ and the near-field effects in a CHISPR protocol.
\section{Chiral Surface Plasmon Resonance (CHISPR)}
\begin{figure}[t!]
		\includegraphics[width=0.7\linewidth]{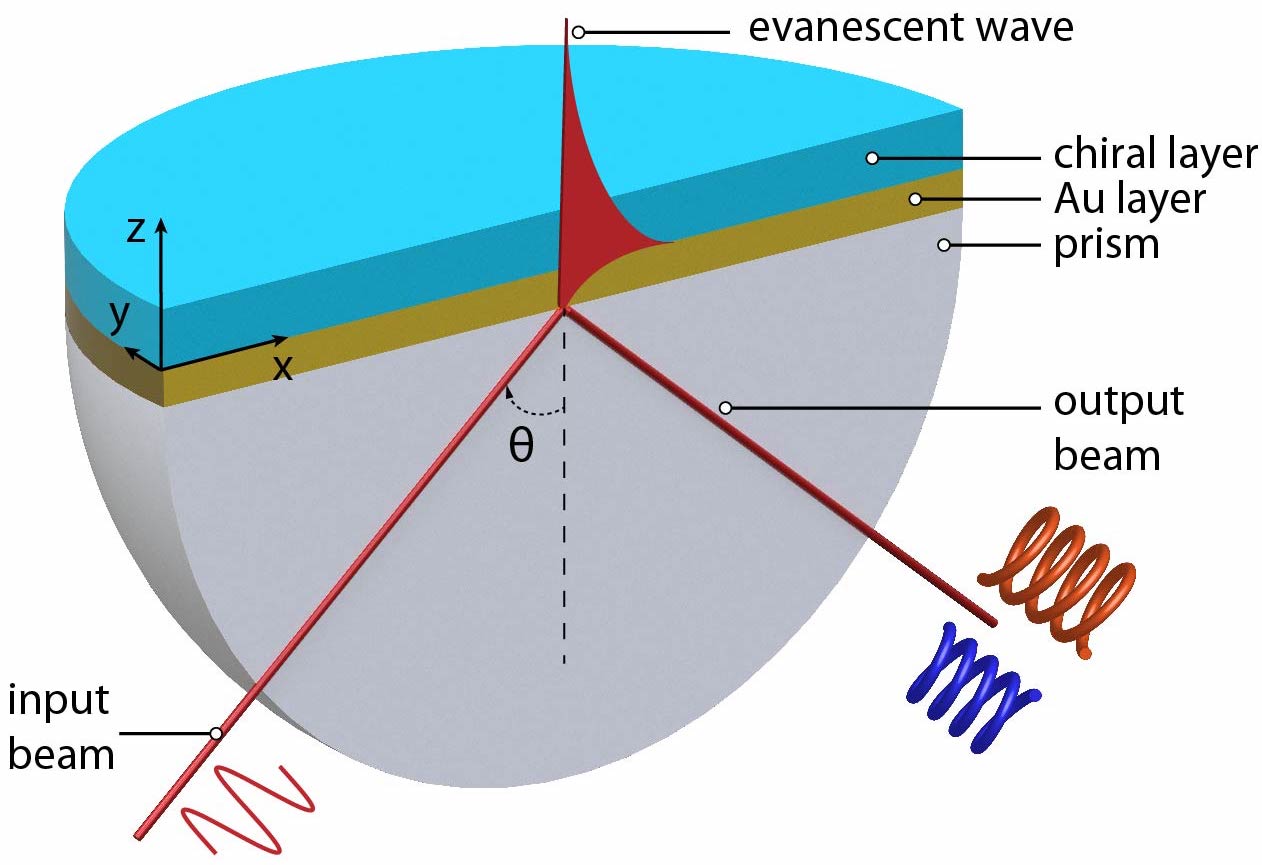}
	\caption{A surface plasmon resonance experimental setup (Kretschmann configuration) for the detection of chirality from thin (sub-wavelength) chiral layers. A linearly TM(p) polarized beam incident on a thin gold layer (Au layer thickness $\sim50$\,nm) excites a surface plasmon polariton (SPP) (indicated by the evanescent wave) at a particular angle, $\theta$, which propagates along the metal-chiral interface. The SPP wave is modified by the chiral environment, resulting in an outgoing optical chirality flux which can be used to infer the properties of the chiral layer (see text for details).}	
    	\label{fig:fig1}
\end{figure}

The refractive index $n_+$($n_-$) for a right- (left-) circularly polarized (RCP/LCP) wave traversing a chiral substance is given by $n_\pm = n_c \pm \kappa$, where $n_c$ is the average refractive index of the chiral substance, and $\kappa$ is the substance's chirality parameter\,\cite{barron_2004}. Traditional ORD and CD are typically transmission measurements, in which the rotation and absorption signals are linearly proportional to the chirality parameter $\kappa$\,\cite{barron_2004}. In particular, rotation is proportional to Re($\kappa$) while absorption to Im($\kappa$). However, if the chiral material thickness becomes comparable to the interrogation wavelength, ORD and CD signals become $<\!1\,\mu$deg\,\cite{Sofikitis2014,Bougas2015} (in the visible range; mdeg CD signals are attainable in the UV, where strong absorption lines are present\,\cite{barron_2004}) and, therefore, impractical to measure (mainly due to background contributions). An alternative configuration for probing the chiral parameter is the measurement of the reflection and/or refraction from a dielectric-chiral interface\,\cite{Silverman1992,Ghosh2006}. In such reflection-based measurements it is possible to enhance the chiral asymmetry effects near the critical angle (which scale as $\sqrt{\kappa}$, with $\kappa \ll 1$)\,\cite{SILVERMAN1989,Sofikitis2014,Bougas2015}, when the material is electrically thick and is probed from the prism side (no critical angle exists from the air side). However, for chiral material thicknesses comparable to the interrogation wavelength, reflection measurements are not possible anymore as the chiral-air interface cannot be ignored and the critical angle is now determined by the prism-air refractive index contrast. Overall, path-length enhancement techniques can allow for enhanced ORD in transmission and reflection, but for lossy systems, as is typically the case for liquid and/or solid samples, these also become inadequate (in the case of CD path-length enhancement techniques can be used, but mainly for weak molecular transitions\,\cite{Mueller2000}). On the other hand, techniques exploiting the enhanced intensity and the superchiral nature of near-fields close to nanophotonic structures, may result in enhanced molecular CD responses, and can be sensitive to monolayers\,\cite{Hendry2010,Tang2011,Zhao2017}. However, in all demonstrated cases, so far only absorption phenomena are discussed and analyzed, and it remains unclear how the real part of the chirality parameter affects the results and whether it is a parameter detectable in these experiments\,\cite{Hendry2010,Tang2011,Davis2013,Karimullah2015,Luo2017,Zhao2017,Mohammadi2018}.\\
\indent To overcome the above-mentioned limitations, we present here an angle-resolved SPR measurement scheme that allows for the absolute measurement of chirality, particularly for the case of thin (sub-wavelength) samples. We consider a SPR setup in the Kretschmann configuration\,\cite{Kret72}, where a metal layer (typically gold) is deposited on a prism surface, upon which the chiral substance is placed. A schematic of the setup is shown in Fig.\,\ref{fig:fig1}.\\
\indent Surface plasmon polariton (SPP) waves are surface electromagnetic excitations confined to and propagating along a metal-dielectric or metal-air interface. To excite the SPP wave, a TM (p)-polarized wave is required (components $E_x$, $H_y$, $E_z$; Fig.\,\ref{fig:fig1}). The incident wave must have a high tangential wavenumber $k_{\rm inc}$ to match the high SPP wavenumber $k_{\rm SPP}$ and achieve efficient power transfer to the SPP wave. In an angle-resolved experiment $k_{\rm inc}$ is controlled both by the angle of incidence, $\theta$ (Fig.\,\ref{fig:fig1}), and the refractive index of the substrate (prism), $n_{\rm sub}$, via $k_{\rm inc} = k_{0}\cdot n_{\rm sub} \cdot \sin(\theta)$, where $k_0$ is the free-space wavenumber\,\cite{Schasfoort2017}. As $\theta$ is scanned, maximum power transfer from the incident wave to the SPP wave is achieved at a certain angle and the excitation of the SPP wave is, therefore, manifested as a dip in an angle-resolved measured reflection. Contrary to the typical case of metal-dielectric interfaces, the presence of the chiral layer qualitatively changes the SPP wave, as was shown in Ref.\,\cite{Mi2014}, generating an s-wave at the metal-chiral interface, and as such, the properties of a chiral layer should be observable through angle-resolved SPR measurements.\\
\begin{figure}[t!]
		\includegraphics[width=0.8\linewidth]{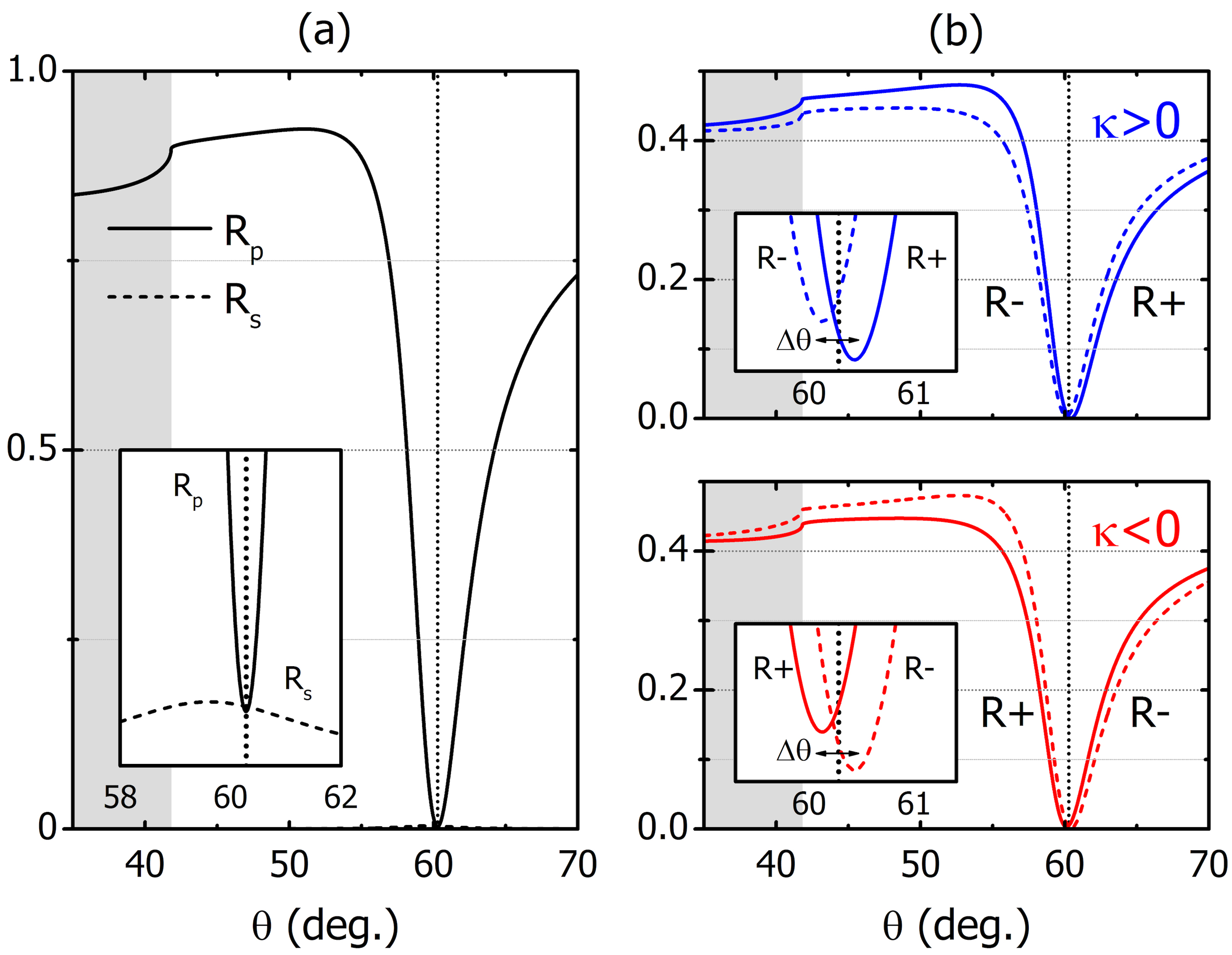}
	\caption{SPR reflectance under the presence of a 100\,nm thin chiral layer with $n_c= 1.33$. The SPR is excited with a $p$-wave and the reflected power is analyzed into (a) $p$- and $s$- components, $R_p$, $R_s$, respectively (same for $\kappa= \pm0.1$), and (b) RCP(+) and LCP(-) components, $R_+$, $R_-$, respectively (top: $\kappa = +0.1$, bottom: $\kappa = -0.1$). The effect of chirality appears in (a) as a chiral-dependent angular shift of $R_p$, accompanied by nonzero $R_s$ and in (b) as a chiral-dependent angular split ($\Delta\theta=\theta_+-\theta_-$) between $R_+$ and $R_-$. The magnitude and sign of $\Delta\theta$ depends on $|\kappa|$ and sgn($\kappa$), respectively. The vertical dashed lines denote the angle of minimum $R_p$, i.e. the SPR angle, and the shaded areas denote the region below the critical angle (41.8\,deg.).}	
    	\label{fig:fig2}
\end{figure}
\indent In Fig.\,\ref{fig:fig2} we present the results of a simulated angle-resolved SPR experiment in the presence of a thin chiral layer. Here, we assume operation at 633\,nm, a typical wavelength for SPR instruments, and consider a prism (refractive index 1.50) coated with a 50\,nm thin Au layer (Au permittivity: $-11.753 + 1.260\,i$ at 633\,nm\,\cite{JohnsonChristy}). On top of the Au layer we place a 100\,nm thin layer of a chiral substance that we assume to be dispersed in water, and therefore we choose an average refractive index $n_{\rm c}\,=\,1.33$. To clarify our findings, we use a large chirality parameter $\kappa$ and consider both possibilities for the sign, i.e. $\kappa = \pm0.1$. In order to excite the SPP wave we send a $p$-polarized wave at angle $\theta$, as illustrated in Fig.\,\ref{fig:fig1}. We analyze the reflected wave in terms of $s$ and $p$ components and calculate the power at each polarization, namely $R_{s}$, $R_{p}$. Additionally, we analyze the total reflected power $R_{s}+R_{p}$ in terms of $+/-$ components, which we denote as $R_{\pm}=|r_{\pm}|^2$, where $r_{+}$\,($r_{-}$) is the complex amplitude of the RCP (LCP) wave (that is, $R_++R_-=R_s+R_p$). In an actual experiment, measurement of $R_{s/p}$ and $R_{+/-}$ can be easily performed with the incorporation of a Stokes polarimeter at the analysis stage.\\
\indent In Fig.\,\ref{fig:fig2}a we show the reflected power measured in terms of $s$/$p$ waves, as is typically performed in SPR experiments. The $R_p$ curve has a pronounced reflection-dip at 60.3\,deg., indicating the excitation of a SPP wave, while we also observe a nonzero $R_s$ peaking at 59.5\,deg. (Fig.\,\ref{fig:fig2}a, inset), as now part of the $p$-wave is transferred to the $s$-wave due to the presence of the chiral layer\,\cite{Mi2014}. We emphasize here, that for $\kappa = 0$, not only $R_s=0$ but also the $R_p$ reflection-dip is located at 60.1\,deg., that is, $\kappa$ induces an angular shift on $R_p$. This shift is identical for both $\kappa = \pm0.1$ and, hence, this measurement cannot differentiate between left-handed and right-handed chiral substances. However, when we analyze the reflected wave in terms of RCP/LCP ($+/-$) components, we observe that the minima of the $R_+$, $R_-$ reflectances do not coincide, but are separated by an angle $\Delta\theta\equiv \theta_+ - \theta_-$, where $\theta_+$ ($\theta_-$) denotes the angle of the $R_+$ ($R_-$) minimum (Fig.\,\ref{fig:fig2}b). Moreover, we observe that for $\kappa>0$ ($\kappa<0$), $\Delta\theta>0$ ($\Delta\theta<0$). Thus, the presence of a thin chiral layer results in a chiral-dependent angular split $\Delta\theta$ between the measured reflectances of $R_+$ and $R_-$, which has a distinct behaviour depending on the sign and magnitude of $\kappa$.\\
\begin{figure}[t!]
		\includegraphics[width=0.9\linewidth]{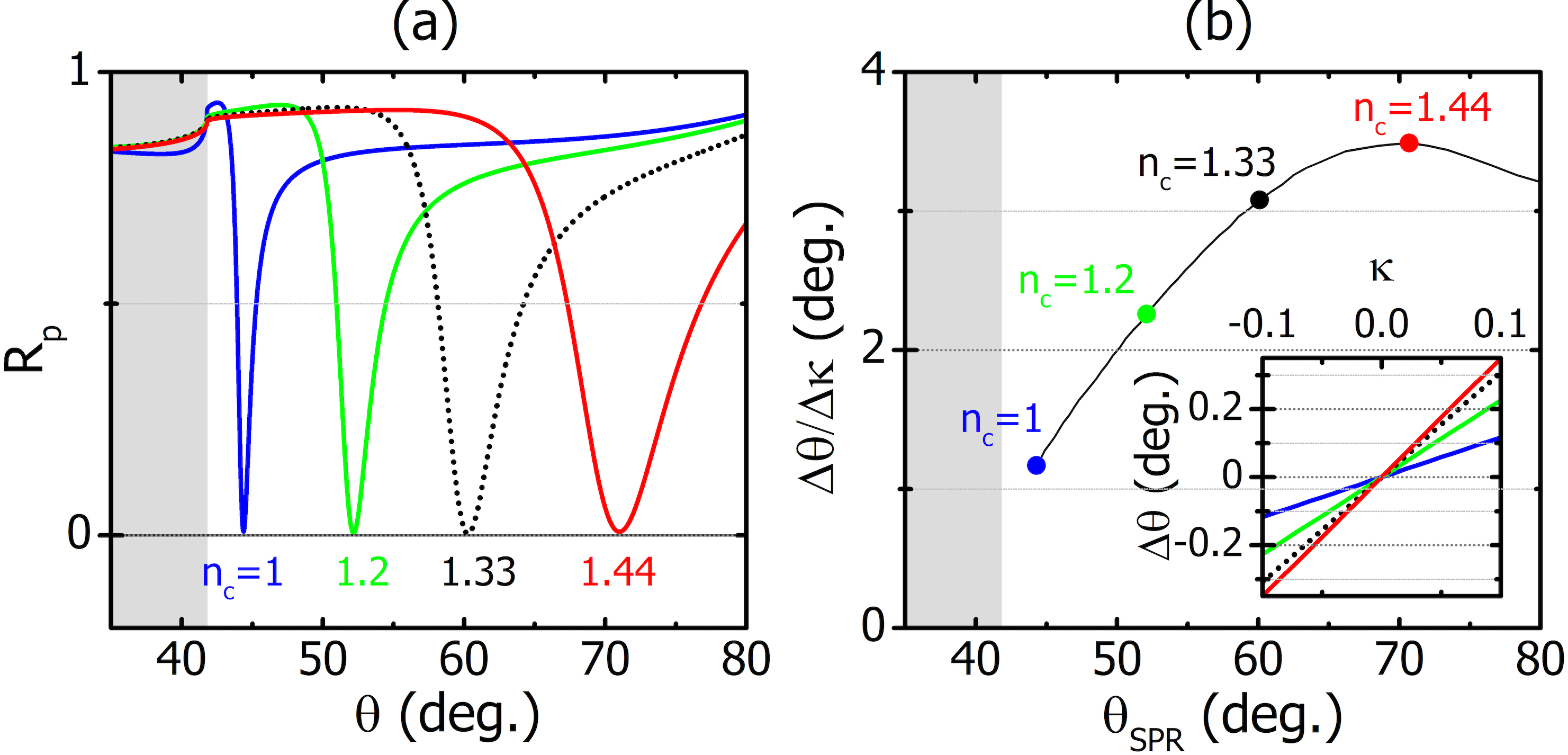}
	\caption{Measurement sensitivity of $\Delta\theta$. (a) Angle-resolved $R_p$ reflectance denoting the SPR angle ($\theta_{\rm{SPR}}$) for the shown selected values of $n_c$ with $\kappa = 0$. (b) $\Delta\theta/\Delta\kappa$ vs SPR angle. The cases for $n_c$ shown in (a) are marked with dots of the same colour. Inset: $\Delta\theta$ vs $\kappa$ for the selected values of the host index $n_c$. The dotted line denotes the system of Fig.\,\ref{fig:fig2} with $n_c = 1.33$.}	
    	\label{fig:fig3}
\end{figure}
\indent In Fig.\,\ref{fig:fig3} we show the dependence of the angular split $\Delta\theta$ on the chiral substance's refractive index, $n_c$, and we focus on the case of real $\kappa$ only (the case of imaginary $\kappa$ is discussed in a following section). We observe (a) a linear dependence between the magnitudes of $\Delta\theta$ and $\kappa$ (Fig.\ref{fig:fig3}b inset), (b) a distinct correspondence between the signs of $\Delta\theta$ and $\kappa$ (Fig.\ref{fig:fig3}b inset), and (c) a non-monotonic dependence of $\Delta\theta$ on $n_c$ (Fig.\ref{fig:fig3}b). This non-monotonic dependence is related to the interplay between the coupling strength of the incident wave to the SPP wave and the interaction strength of the SPP wave with the chiral layer. In particular, as $n_c$ increases, the SPR dispersion changes and the reflection-dip is shifted to higher angles due to higher $k_{\rm SPP}$ (Fig.\,\ref{fig:fig3}a). In turn, the coupling of the incident wave to the SPP becomes stronger, leading to higher $\Delta\theta$ and, hence, in increased sensitivity (Fig.\,\ref{fig:fig3}b). Eventually, for very high incident angles the coupling of the incident wave to the SPP becomes weaker, leading to weaker $\Delta\theta/\Delta\kappa$ accordingly. For angles close to the critical angle, the effect becomes the weakest, verifying that the measurement is mediated entirely by the SPP wave and is not associated with the total-internal-reflection angle, as in the case of reflection from a dielectric-chiral interface\,\cite{SILVERMAN1989}. Thus, by measuring the magnitude and sign of this chiral-dependent angular split, we obtain absolute information about the magnitude and sign of $\kappa$.\\
\indent To fully understand the mechanism behind the chiral-dependent SPR-reflectance angular split, we examine how the near-field properties of the SPP wave are associated with the properties of the reflected wave in the far-field. We start by analyzing the SPP wave along its propagation direction ($x$) into $+/-$ components, i.e. $\mathbf{A}_{\rm SPP}=A_y \hat{y}+A_z \hat{z}= A^{+}_{\rm SPP}(\hat{y}+i\hat{z})+A^{-}_{\rm SPP}(\hat{y}-i\hat{z})$, where $A^{\pm}_{\rm SPP}=(A_y\mp A_z)/2$ and $\mathbf{A}$ is any of the electromagnetic field quantities $\bf{E}$, $\bf{H}$, $\bf{B}$, $\bf{D}$; then, we calculate the electric and magnetic energy densities $w^{\pm}_e=(1/4)\bf{E}^{\pm}_{\rm SPP}(\bf{D}^{\pm}_{\rm SPP})^*$ and $w^{\pm}_m=(1/4)\bf{B}^{\pm}_{\rm SPP}(\bf{H}^{\pm}_{\rm SPP})^*$, respectively, which we integrate to find the total energy stored in each of the two ($+/-$) components, namely $W_{\pm}=\int_{\rm{V}}(w^{\pm}_e+w^{\pm}_m)\,d^3x$ (Fig.\,\ref{fig:fig4}a). Here, the integration volume V is the entire SPP volume extending above the metal (where the chiral layer is to be probed). For $\kappa=0$ we obtain $W_+=W_-$, as the SPP wave has only an $E_z$-component on the $yz$-plane, which is equally distributed between the two $+/-$ components (typical nonchiral SPR case). This is shown in Fig.\,\ref{fig:fig4}b, where the energy difference $W_+-W_-$ is normalized to the incident energy $S_{\rm inc}/2\omega$ ($\omega$ is the angular frequency and $S_{\rm inc}$ is the magnitude of the time-averaged Poynting vector). However, the onset of chirality causes the emergence of an $E_y$-component\,\cite{Mi2014} and, hence, an unbalanced storage of the optical energy between the $+/-$ components of the SPP. In fact, for $\kappa>0$ ($\kappa<0$), RCP (LCP) components are favoured and, therefore, $W_+>W_-$ ($W_+<W_-$ ) (Fig.\,\ref{fig:fig4}b). This stored energy excess between $+/-$ SPP components in the near-field results in nonzero $R_s$ reflectance in the far-field; this is apparent in the fact that the peak of $R_s$ coincides with the peak of $W_+-W_-$ at 59.5\,deg. which differs from the $R_p$ minimum at 60.3\,deg. (Fig.\,\ref{fig:fig2}a and Fig.\,\ref{fig:fig4}b). In other words, the $E_y$-component that emerges in the near-field due to chirality, is identified in the far-field as well, as power transfer from the outgoing $p$-wave to the $s$-wave. When the total reflected wave in the far-field is analyzed in $+/-$ components as well, the reflectance splits into two parts, which have their minima at different angles (Fig.\,\ref{fig:fig2}b). This angular split is mediated by the resonance of the surface plasmon; the amplitude of the $E_y/E_x$ ratio is symmetric around the SPR angle (Fig.\,\ref{fig:fig4}c), however, the phase arg$(E_y/E_x)$ undergoes a $\pi$-shift, favouring the advance of either the $E_x$ or the $E_y$ component, depending on whether the angle of incidence is below or above the SPR angle (Fig.\,\ref{fig:fig4}c). Consequently, the mixture of the reflected RCP and LCP wave-components is weighted differently, resulting in an excess of either RCP or LCP waves below or above the SPR angle and, hence, a reflectance split between $R_+$ and $R_-$ waves. As for the magnitude of $\kappa$, it does not significantly affect the arg($E_y/E_x$) (it induces a slight angular shift), however, it notably changes the amplitude of $E_y/E_x$, which increases with increasing $\kappa$. Changing the sign of $\kappa$ induces a $\pi$-shift of the $E_y$ phase, without affecting $E_x$ (Fig.\,\ref{fig:fig4}c). Hence, the sign of $\kappa$ does not affect the amplitude of the ratio $E_y/E_x$ but causes the interchange between RCP/LCP components.
\begin{figure}[t!]
		\includegraphics[width=0.9\linewidth]{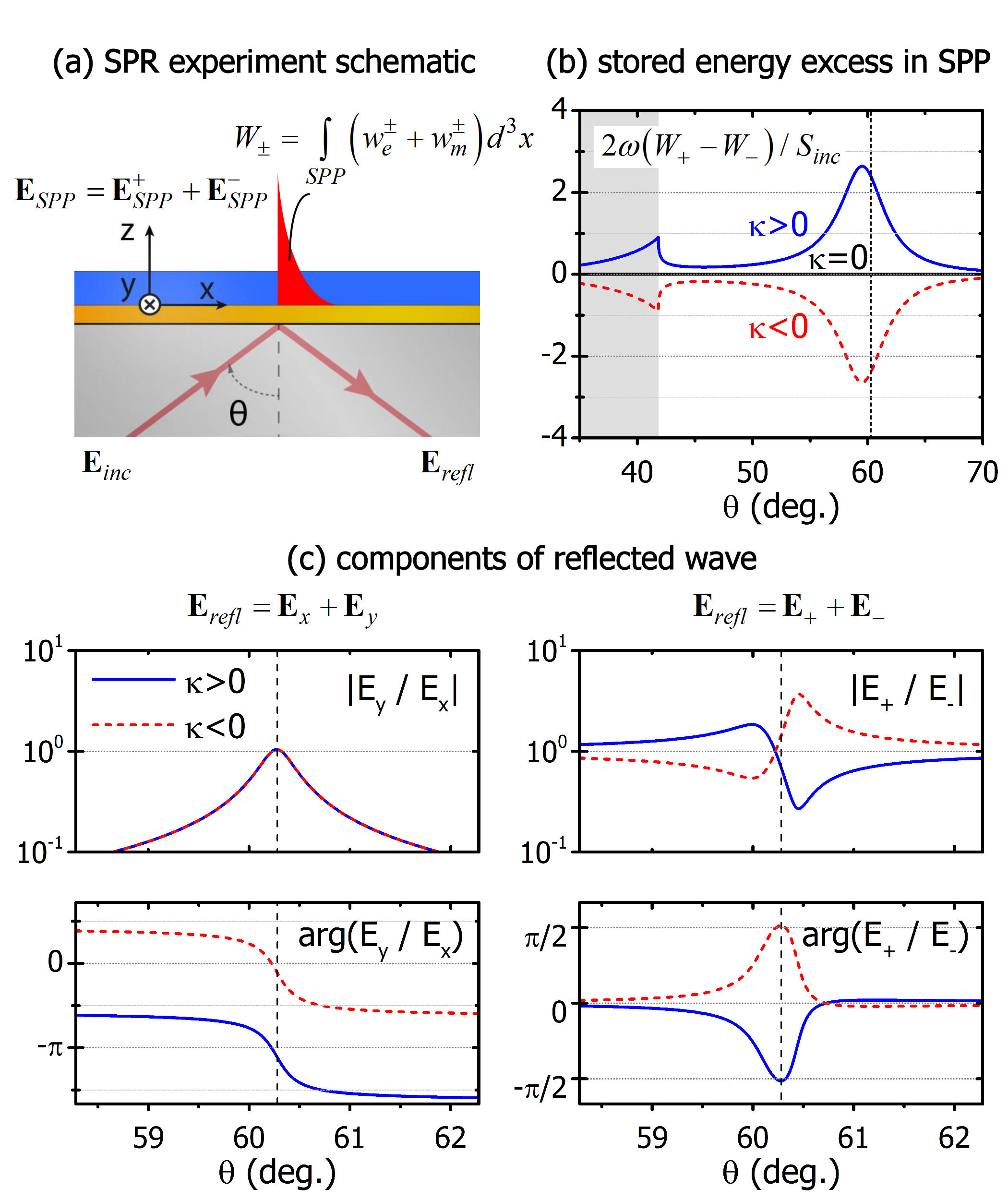}
	\caption{Mechanism of the $R_+$, $R_-$ angular split. (a) The incident wave excites the SPP, which we analyze in RCP/LCP ($+/-$) components along its propagation direction. The total energy stored in each component is denoted as $W_\pm$ (see also text for details). (b) Stored energy difference  between RCP/LCP components of the SPP wave (normalized over the incident energy $S_{\rm inc}/2\omega$). Chirality causes energy excess between RCP/LCP waves, which changes sign upon sign change of $\kappa$ ($\kappa=\pm0.1$). (c) Ratio of amplitude and phase of the reflected wave components for $\kappa=+0.1$ (solid blue lines), $\kappa=-0.1$ (dashed red lines). We analyze the wave in terms of $s/p$ components (left) and RCP/LCP components (right). We present the amplitude in logarithmic scale to emphasize the inversion symmetry of the ratio $|E_+/E_-|$ upon sign change in $\kappa$. The vertical dashed line denotes the angle of minimum $R_p$, i.e. the SPR angle.}
    	\label{fig:fig4}
\end{figure}
\section{Differential measurements}
We consider now measurement configurations based on differential signals which are immune to signal fluctuations and drifts. Specifically, we consider two relevant quantities associated with the reflected (outgoing) RCP/LCP waves: the amplitude and phase differential reflectance (DR), namely $\rho_{\rm{\small DR}}$ and $\phi_{\rm{\small DR}}$, respectively. We define the DR signals as,
\begin{equation}
\rho_{\rm{\small DR}}=\frac{|r_+|^2-|r_-|^2}{|r_+|^2+|r_-|^2}\,, \,\,\rm{and} \,\,\phi_{\rm{\small{DR}}}=\rm{Arg}\Big[\frac{r_+}{r_-}\Big]\,.
\end{equation}
\begin{figure}[t!]
		\includegraphics[width=0.9\linewidth]{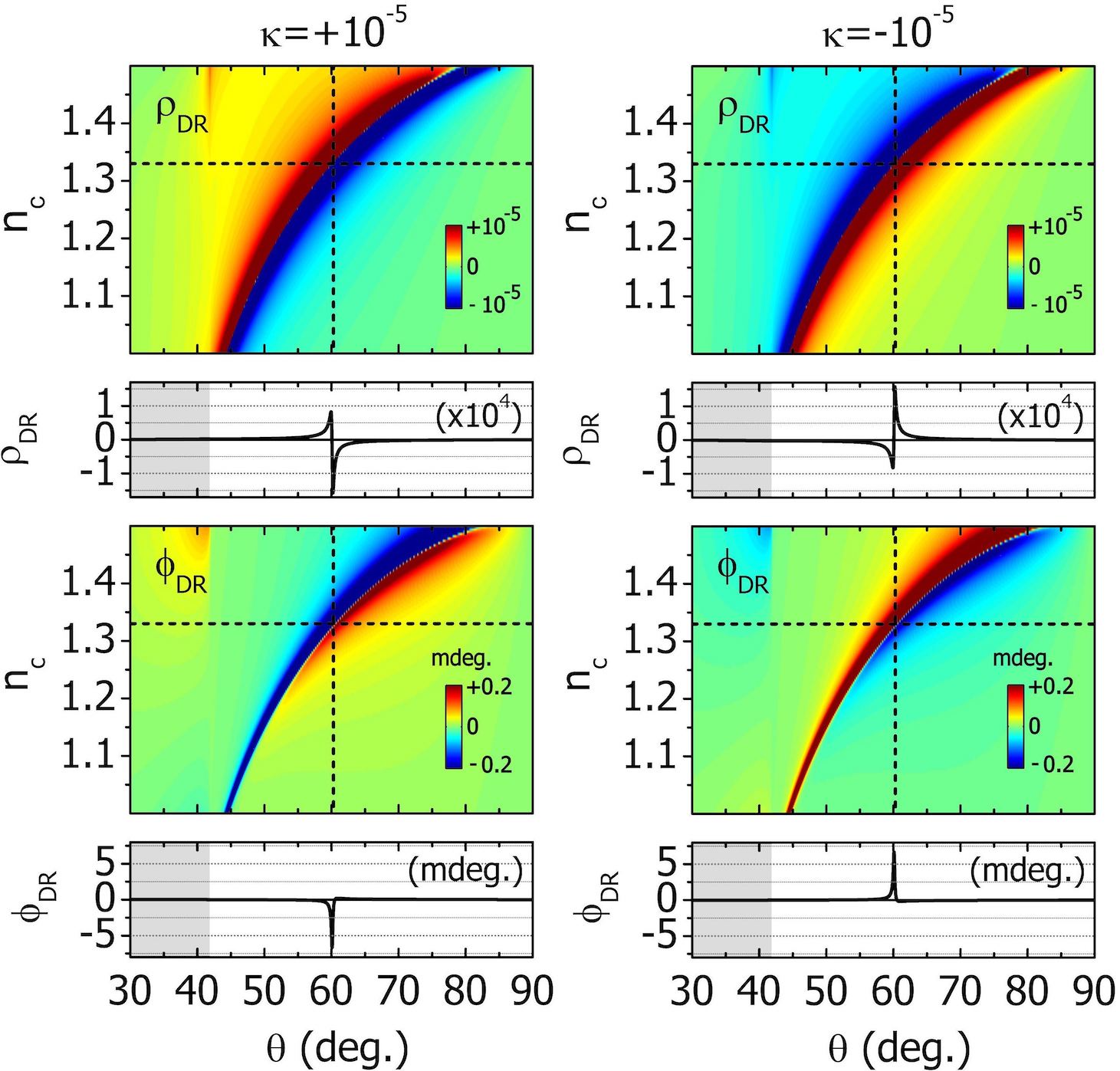}
	\caption{Differential reflectance (DR) signals for a 100\,nm thin chiral layer with $\kappa = \pm10^{-5}$, as a function of the background index of the chiral layer ($n_c$). Specific examples for $n_c = 1.33$ are marked with horizontal dashed lines and are shown separately below each panel. The vertical dashed line marks the SPR angle for the chosen value of $n_c$.}
    	\label{fig:fig5}
\end{figure}
In Fig.\,\ref{fig:fig5} we present the DR signals, $\rho_{\rm{\small DR}}$ and $\phi_{\rm{\small DR}}$, respectively, for a value of $\kappa = \pm10^{-5}$ (which is a realistic value for the chirality parameter for, e.g., aqueous solutions of chiral molecules\,\cite{Sofikitis2014,Bougas2015}), as a function of the background index of the chiral layer, $n_c$. We observe $\rho_{\rm{\small DR}}$ signals of the order of $\sim10^{-4}$ and $\phi_{\rm{\small DR}}$ signals of the order of a few $\sim$mdegs, both within feasible sensitivity of SPR instruments\,\cite{Piliarik2009,Wang2011,Pooser2016}. As a comparison, we note that the optical rotation signal from a transmission measurement of a 100\,nm chiral layer with $\kappa=+10^{-5}$ at 633\,nm, is $\sim3\times10^{-4}$\,deg.\,\cite{barron_2004}. Moreover, we observe that $\rho_{\rm{\small DR}}(-\kappa) = -\rho_{\rm{\small DR}}(\kappa)$ and $\phi_{\rm{\small DR}}(-\kappa) = -\phi_{\rm{\small DR}}(\kappa)$. Thus, with this type of measurement we are able to quantify Re($\kappa$) (magnitude and sign) with increased sensitivity compared to measurements of $\Delta\theta$\,\cite{Kabashin2009,Patskovsky2010,Wang2011}.\\
\indent Additionally, we observe that both differential signals, $\rho_{\rm{\small DR}}$ and $\phi_{\rm{\small DR}}$, decrease in amplitude as the SPR moves away from the critical angle, contrary to $\Delta\theta/\Delta\kappa$ which we observe to increase (Fig.\ref{fig:fig3}). This decrease is related to the broadening of the SPR feature due to increased losses for higher $k_{\rm SPP}$, and to the reduction of the $R_s/R_p$ ratio, which expresses the strength of the $p$- to $s$-wave conversion. In Fig.\,\ref{fig:fig6} we show the change in the $R_s/R_p$ ratio with increasing SPR angle (due to increasing $n_c$). We observe that the $R_s/R_p$ ratio decreases while simultaneously broadening, which yields, thus, reduced differential signals. Moreover, the peak-to-peak values of $\rho_{\rm{\small DR}}$ and $\phi_{\rm{\small DR}}$  ($\Delta\rho_{\rm{\small DR}}$ and $\Delta\phi_{\rm{\small DR}}$, respectively; Fig.\,\ref{fig:fig6}b), qualitatively follow a similar trend indicating a strong connection with the strength of $R_s/R_p$. Furthermore, we observe that the variation of $R_s/R_p$ (and consequently of $\Delta\rho_{\rm{\small DR}}$ and $\Delta\phi_{\rm{\small DR}}$) is non-monotonic and it generally depends on the properties of the particular metal. However, we emphasize that regardless of the exact variation, $\rho_{\rm{\small DR}}$ and $\phi_{\rm{\small DR}}$ allow for unambiguous determination of $\kappa$.
\begin{figure}[t!]
		\includegraphics[width=0.8\linewidth]{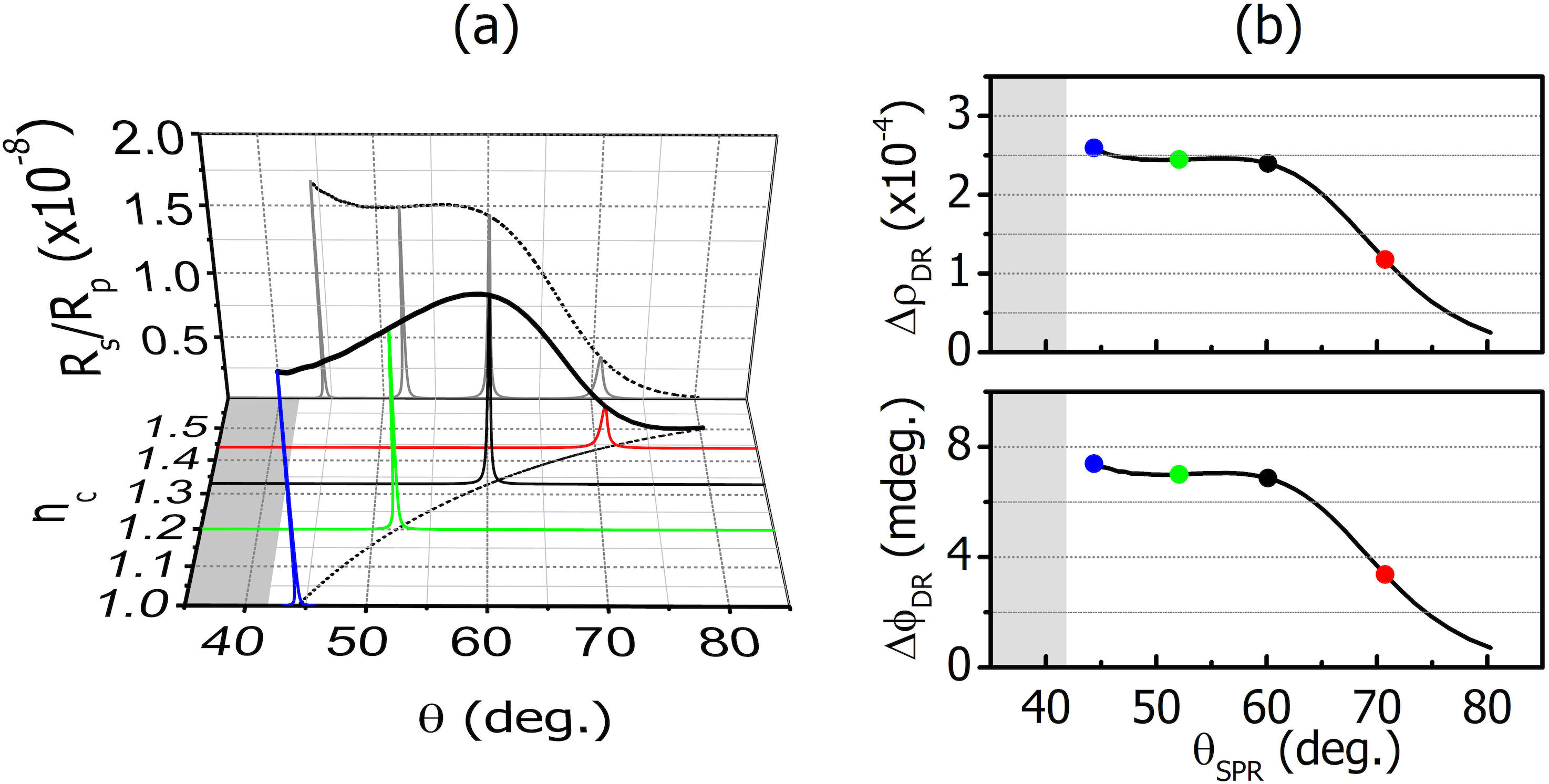}
	\caption{Effect of coupling strength between $s-$ and $p-$waves due to chirality on differential reflection measurements. (a) $R_s/R_p$ ratio as function of $n_c$ for $\kappa = \pm10^{-5}$. (b) Peak-to-peak values of the differential signals $\rho_{\rm DR}$ and $\phi_{\rm DR}$ [$\Delta\rho_{\rm DR}$ (top) and $\Delta\phi_{\rm DR}$ (bottom), respectively]. The marked cases in (a), (b) correspond to the cases shown in Fig.\,\ref{fig:fig3} using the same colour-code.
	}	
    	\label{fig:fig6}
\end{figure}
\section{Absorption effects}
\begin{figure}[t!]
		\includegraphics[width=0.9\linewidth]{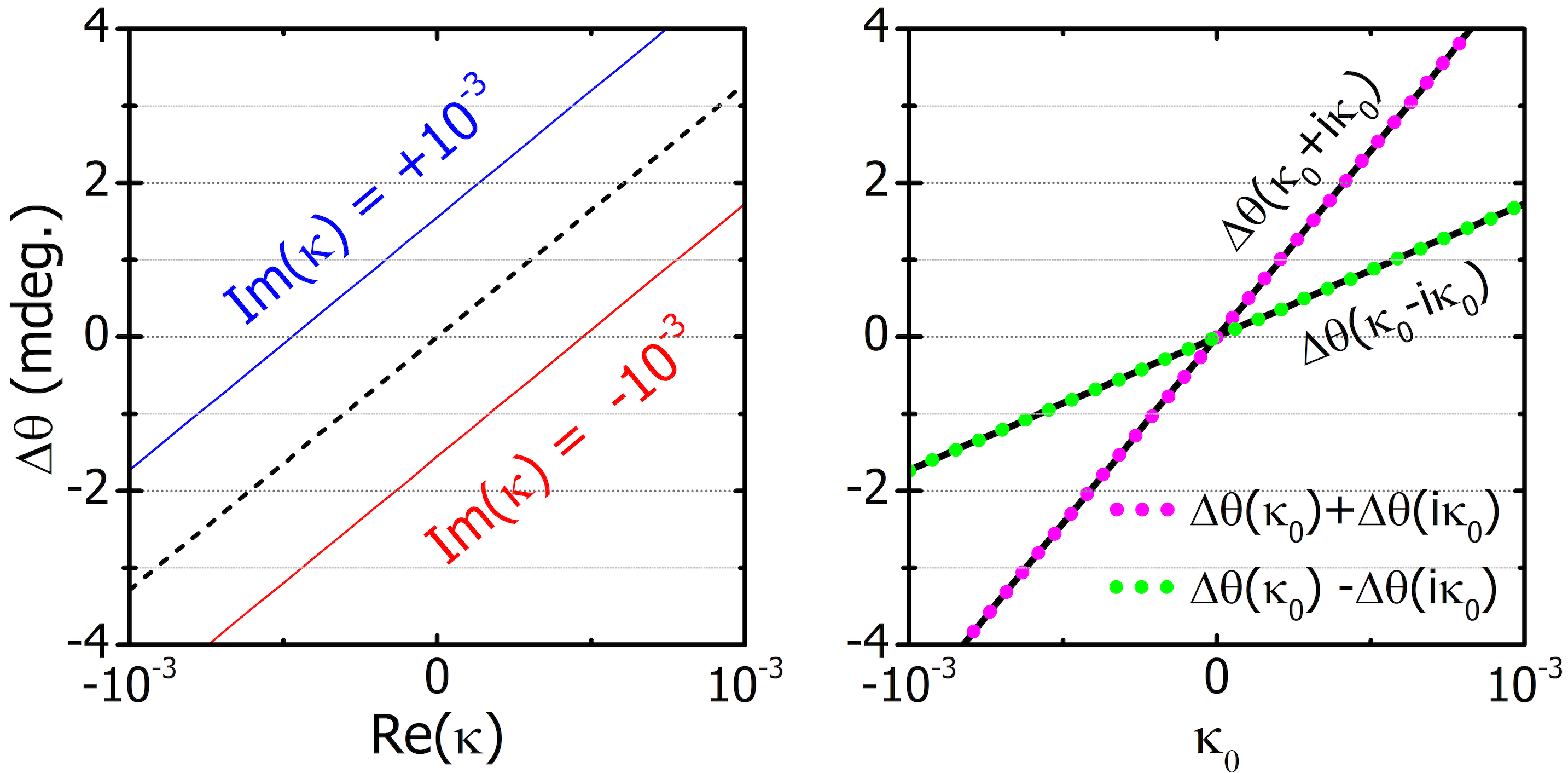}
	\caption{Measurement sensitivity of $\Delta\theta$ in the presence of molecular absorption, i.e. Im($\kappa$)$\neq0$. (a) Scan of Re($\kappa$) for Im($\kappa$)$=\pm 0.001i$ [dashed line corresponds to Im($\kappa$)=0]. (b) Demonstration of linearity of the effects of Re($\kappa$) and Im($\kappa$) on $\Delta\theta$. Here, we show $\Delta\theta$ for $\kappa = \kappa_0$ and $\kappa = i\,\kappa_0$ separately, and for $\kappa = \kappa_0\pm i\,\kappa_0$. To maintain the passivity of the system, we add artificial loss to the chiral layer index, which now is $n_c = 1.33 + 0.01i$.}	
    	\label{fig:fig7}
\end{figure}

\begin{figure}[t!]
		\includegraphics[width=0.9\linewidth]{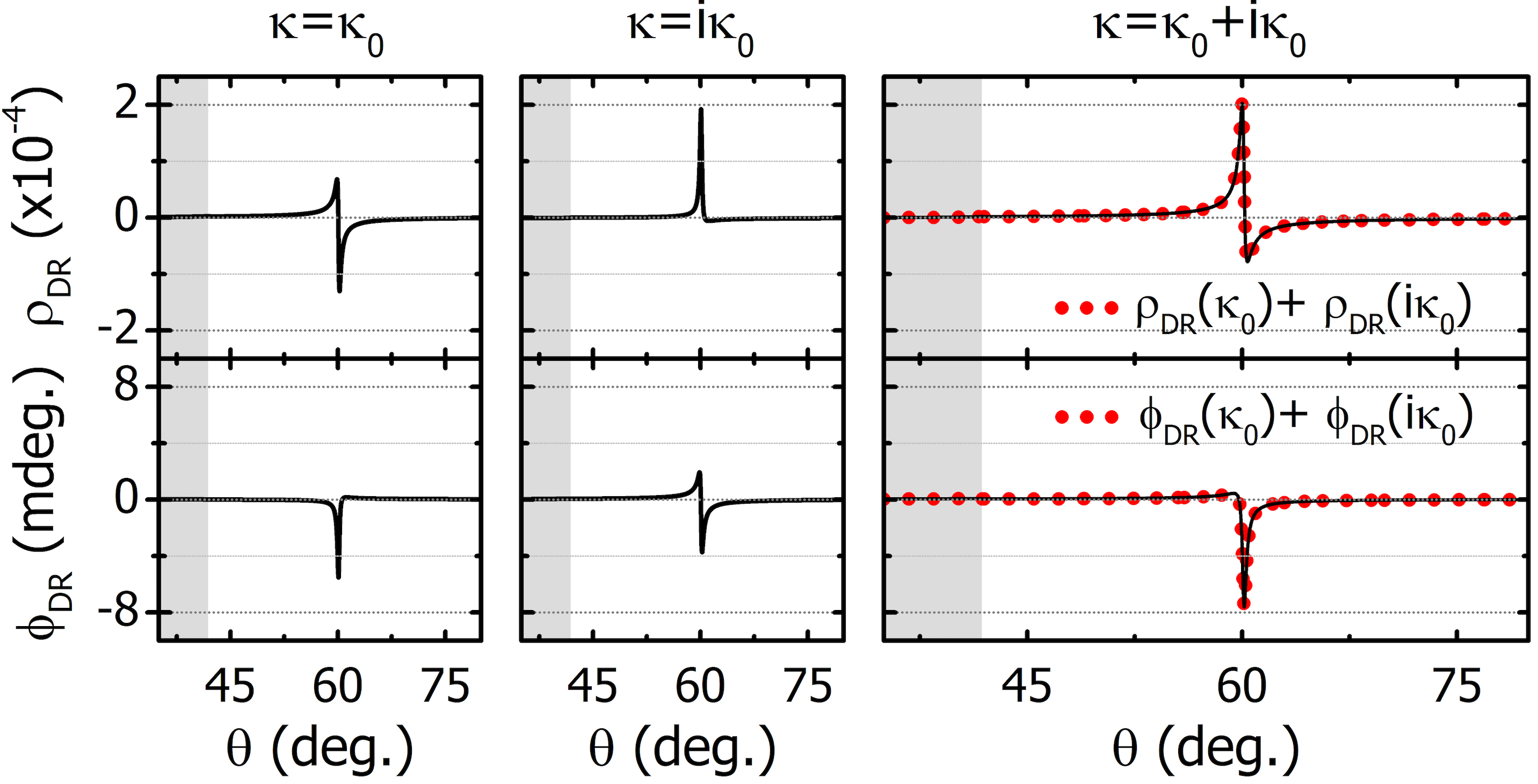}
	\caption{Demonstration of linearity of the effects of Re($\kappa$) and Im($\kappa$) on the differential signals $\rho_{\rm{DR}}$ and $\phi_{\rm{DR}}$. The calculations have been performed for $\kappa = \kappa_0$ (left) and $\kappa = i\,\kappa_0$ (middle) separately, and for $\kappa = \kappa_0+i\,\kappa_0$ (right), with $\kappa_0 =10^{-5}$. Here we use: $n_c = 1.33 + 10^{-3}i$.}
    	\label{fig:fig8}
\end{figure}

In the previous sections we consider only real-valued chirality parameter $\kappa$. This is the case when one performs measurements at optical frequencies far detuned from any molecular resonances, and, as such, absorption - which is proportional to Im$(\kappa)$ - is negligible\,\cite{barron_2004}. When the optical frequency of an SPR instrument is near a molecular transition, absorption becomes substantial and the proposed measurements should be interpreted with care. In order to introduce an imaginary part in the chirality parameter $\kappa$ without violating the passivity we must ensure that Im$(n_c\pm \kappa)>0$. Hence, we introduce artificial loss in the average refractive index of the chiral layer, which now is $n_c = 1.33 + 0.01\,i$. \\
\indent In Fig.\,\ref{fig:fig7} we present $\Delta\theta$ as a function of Re$(\kappa)$ for Im$(\kappa) = -10^{-3},0, 10^{-3}$. In the presence of absorption (Im$(\kappa)\neq0$) the angular split $\Delta\theta$ obtains a linear chiral-dependent offset, and the effects of Re($\kappa$) and Im($\kappa$) appear as linear superpositions in the total $\Delta\theta$. As an example, in Fig.\,\ref{fig:fig7} we use $\kappa_0$ (real variable) to control the strength of the real and imaginary part of $\kappa$ and we calculate $\Delta\theta$ for $\kappa = \kappa_0$, $\kappa = i\,\kappa_0$ and $\kappa = \kappa_0\pm i\,\kappa_0$. The plots of $\Delta\theta(\kappa_0 \pm i \kappa_0)$ and $\Delta\theta(\kappa_0)\pm \Delta\theta(i\,\kappa_0)$ are identical and, thus, verify that the effects of Re($\kappa$) and Im($\kappa$) are linearly superimposed, i.e. cumulative. \\
\indent In Fig.\,\ref{fig:fig8} we show the change in the differential signals, $\rho_{\rm{\small DR}}$ and $\phi_{\rm{\small DR}}$, in the presence of absorption. Here, we again assume a $\kappa_0$ (real variable), and we calculate $\rho_{\rm{\small DR}}$ and $\phi_{\rm{\small DR}}$ for $\kappa = \kappa_0$ and $\kappa = i\,\kappa_0$, and also the sum of the two signals. Additionally, we present the same quantities for $\kappa = \kappa_0+i\,\kappa_0$, which we show to coincide with the sums of the individual signals (Fig.\,\ref{fig:fig8}). 
\section{Optical chirality conservation}
\begin{figure}[ht]
		\includegraphics[width=0.9\linewidth]{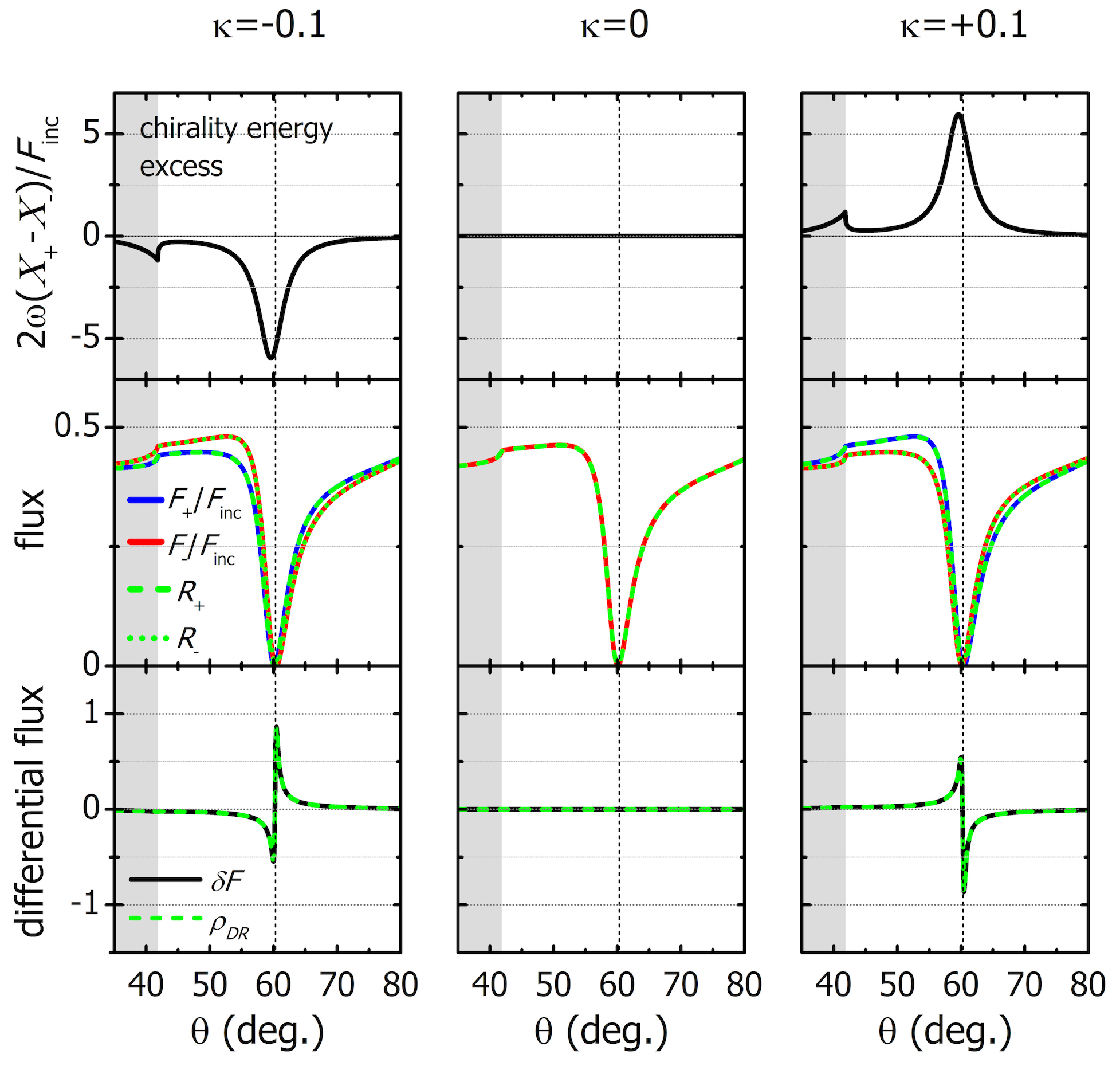}
	\caption{Top row: Stored chirality energy difference $X_+-X_-$ between RCP/LCP ($+/-$) components of the SPP wave, normalized with the incident chirality energy $F_{\rm inc}/2\omega$. Middle row: Reflected chirality flux $F_{\pm}$ (solid blue/red lines), normalized with the incident chirality flux $F_{\rm inc}$, and reflectance $R_{\pm}$ (dashed/dotted lines). Bottom row: differential flux $\delta F$ and differential reflectance amplitude $\rho_{\rm{DR}}$. For these simulations we use a 100\,nm thin chiral layer with $n_c = 1.33$ and $\kappa = -0.1$ (left column), $\kappa = 0$ (middle column) and $\kappa = +0.1$ (right column). In all panels, the SPR angle (angle of minimum $R_p$) is marked with a vertical dashed line.}	
    	\label{fig:fig9}
\end{figure}

The association of the near-field features of the SPP wave with the far-field properties of the reflected wave, that we demonstrated using Poynting's theorem for conservation of energy, is closely related to another conservation law, that of optical chirality density\,\cite{Lipkin1964,Tang2010,Bliokh2011,Coles2012,Philbin2013,Poulikakos2016}. In the time-averaged, time-harmonic case this is written as,
\begin{equation}
-2\omega\int\limits_{\mathrm{V}}\mathrm{Im}(\chi_e-\chi_m)\,d^3x+\int\limits_{\mathrm{V}}\mathrm{Re}(\nabla\cdot\mathbf{F})\,d^3x =0,
\end{equation}
where $\chi_e$ and $\chi_m$ are the electric and magnetic optical chirality densities, respectively, and $\mathbf{F}$ is the corresponding chirality flux:
\begin{eqnarray}
\chi_e=\frac{1}{8} \Big[ \mathbf{D}^* \cdot (\nabla\times \mathbf{E})+\mathbf{E} \cdot (\nabla\times \mathbf{D}^*) \Big],\\
\chi_m=\frac{1}{8} \Big[ \mathbf{H}^* \cdot (\nabla\times \mathbf{B})+\mathbf{B} \cdot (\nabla\times \mathbf{H}^*) \Big],\\
\mathbf{F}=\frac{1}{4} \Big[ \mathbf{E} \times (\nabla\times \mathbf{H}^*)-\mathbf{H}^* \times (\nabla\times \mathbf{E}) \Big].
\end{eqnarray}
For $\kappa\neq0$ the total chirality energy $X$, which is the integral of the total chirality density $\chi=\chi_e+\chi_m$ across the SPP volume, is unequally stored between the $+/-$ SPP components, i.e. $X_+ \neq X_-$, where $X_{\pm}=\int_{\rm{V}}(\chi^{\pm}_e+\chi^{\pm}_m)\,d^3x$. In Fig.\,\ref{fig:fig9} we plot the chirality-energy excess $X_+ - X_-$ for $\kappa= 0,\,\pm0.1$. The result is qualitatively similar to $W_+ - W_-$, as shown in Fig.\,\ref{fig:fig4}b. Due to the chirality conservation law (Eq.\,2), this unbalance results in a chirality flux $\mathbf{F}$ in the far-field, manifested as unequal RCP and LCP components and observed through the angular split or the DR signals. In fact, as shown in Ref.\,\cite{Poulikakos2016} the chirality flux $\mathbf{F}$ of a certain propagating wave is proportional to its power flux $\mathbf{S}=\frac{1}{2} (\mathbf{E} \times \mathbf{H}^*)$ and, in particular, $F_\pm = \mp(\omega/c)S_\pm$ (where $c$ is the vacuum speed of light, and $F_\pm$ and $S_\pm$ are the magnitudes of $\mathbf{F}$ and $\mathbf{S}$ with the signs $+/-$ corresponding to RCP/LCP waves, respectively). As a result, the reflected chirality flux $F_\pm$, normalized by the incident optical power $S_{\rm inc}$, relates directly to the $R_\pm$ power flux (reflectance) as $F_\pm/S_{\rm inc}= \mp(\omega/c)R_\pm$ or $F_\pm/F_{\rm inc} = R_\pm$, where $F_{\rm inc}$ is the incident chirality flux magnitude (see Supporting Information for the definition of $F_{\rm{inc}}$). In Fig.\,\ref{fig:fig9} we present the fluxes $F_\pm/F_{\rm{inc}}$ and $R_\pm$ to emphasize the equivalence between the two quantities. Therefore, we see that our proposed measurement scheme results in a direct measurement of the optical chirality flux, which is directly connected with the near-field optical chirality density\,\cite{Lipkin1964,Tang2010,Bliokh2011,Coles2012,Philbin2013,Poulikakos2016}. Moreover, due to the linear connection between the quantities $F$ and $R$, the differential reflectance amplitude $\rho_{\rm DR}$ is equal to the differential flux $\delta F=(F_+-F_-)/(F_++F_-)$ (see Supporting Information for proof). In Fig.\,\ref{fig:fig9} we also demonstrate the equality between the differential flux $\delta F$ and $\rho_{\rm DR}$.
\section{Discussion and Conclusions}
Plasmonic fields at metal-dielectric interfaces involve field-components with vanishing chirality density and chirality flow that is orthogonal to the propagation direction\,\cite{Bliokh2012}. Intuitively, these properties lead to the conclusion that enhancing the optical chirality by SPPs cannot be possible. This is indeed the case when considering dielectric-metal interfaces. In contrast, at a metal-chiral interface a near-field wave is generated with non-vanishing optical chirality density and chirality flow. In this work, we demonstrate how angle-resolved CHISPR measurements allow for the detection of this near-field wave at the metal-chiral interface, resulting in absolute determination of the chirality of the chiral sample (handedness and magnitude). Moreover, in a CHISPR measurement scheme the whole evanescent-wave volume is sensitive to the probed chiral substance (due to the mobility of the propagating SPPs), contrary to schemes based on local surface plasmons, where the sign and magnitude of the chiroptical response possess a complex dependence on sample geometry \,\cite{Hendry2010,Tang2011,Davis2013,Karimullah2015,Luo2017,Zhao2017,Tang2013}. We demonstrate how a CHISPR scheme is particularly suitable for chiral sensing of thin (sub-wavelength) chiral samples, which are difficult to measure using traditional polarimetric techniques. Furthermore, the CHISPR signals we present are within the sensitivity of current SPR instrumentation for realistic values of the chirality parameter. However, the observed relationships between $\kappa$ and the measured quantities ($\Delta\theta$, $\rho_{\rm DR}$ and $\phi_{\rm DR}$) impose a lower limit of chiral detection. This can be improved by enhancing the local fields, for example via modification of the thin metal layer, e.g. via perforation, in order to take advantage of the strong evanescent fields at the gaps. In future works we will investigate how alternative plasmonic structures can enable enhanced CHISPR detection signals.\\
\indent In conclusion, we show a surface plasmon-aided scheme for probing thin (sub-wavelength) chiral films. Our proposed setup allows for combined measurements to unambiguously identify the sign and magnitude of both real and imaginary parts of the chirality parameter $\kappa$. These findings can be of great interest in chiral-biosensing applications, considering that an angle-resolved CHISPR sensing scheme is a surface-sensitive measurement and differs from conventional chiral-sensing techniques based on transmission measurements. In addition, among the advantages of our scheme are the simplicity of the setup and measurements, as it can be implemented with slight modifications of existing SPR measurement instruments, and has also the potential for miniaturization and portable design\,\cite{Liu2015}. Such a possibility could enable compact devices for real-time sensing of biological processes that occur in limited regions of space.

\section*{Acknowledgements}
The authors thank E. Bernikola, and A. Palmer for their feedback during the completion of this work. In addition, LB thanks G. Iwata and M. Williams for fruitful discussions. This research was supported by European Commission Horizon 2020 (grant no. FETOPEN-737071), ULTRACHIRAL Project. 

\bibliography{CHISPRbib}

\appendix
\renewcommand\thefigure{A\arabic{figure}}  
\renewcommand\thesubsection{A\arabic{subsection}}  
\setcounter{figure}{0}  

\section*{Appendix}
\subsection{Theoretical calculations}
To analytically study our proposed system, we solve Maxwell's equations $\nabla \times \mathbf{E} =i \omega \mathbf{B}$ and $\nabla\times \mathbf{H} = -i \omega \mathbf{D}$ with the appropriate boundary conditions for a chiral layer placed on top of a metal layer, both of which have finite width along the $z$-direction and are infinite on the $xy$-plane (Figure\,1 of the main text). The constitutive relations in the chiral layer region are formulated according to Condon's convention\,\footnote{Condon, E. U. Rev. Mod. Phys. $\mathbf{1937}$, $\mathit{9}$, 432.} as: $\mathbf{D} = \epsilon \epsilon_0 \mathbf{E}+i\,(\kappa/c)\mathbf{H}$ and $\mathbf{B} = \mu \mu_0 \mathbf{H}-i\,(\kappa/c)\mathbf{E}$, where $\epsilon$, $\mu$ refer to the relative permittivity and permeability ($n_c = \sqrt{\epsilon\mu}$), respectively, $\kappa$ is the chirality parameter, and $c$ the vacuum speed of light. By applying the boundary conditions at each of the three interfaces, the tangential $E$- and $H$- components ($E_x$, $E_y$, $H_x$, $H_y$) form a $12\times12$ linear system of equations, which is then solved analytically. To verify our analytical findings we also solve the same problem numerically with full-wave vectorial Finite Element Method (FEM) simulations, utilizing the commercial software COMSOL Multiphysics. The constitutive relations are modified to include the chirality parameter $\kappa$ and the computational space is terminated by Perfectly Matched Layer (PML) sufficiently far from the metal-chiral system.

\subsection{Optical flux, chiral flux, and reflectances}
Analyzing the incident wave in $+/-$ components we calculate the magnitudes of the power flux $S^{\pm}_{\rm{inc}}$ and chiral flux $F^{\pm}_{\rm{inc}}$ for each component. Because the incident wave is linearly polarized (p-wave), we find that these quantities are equally distributed between the $+/-$ components, i.e. $S^{+}_{\rm{inc}}=S^{-}_{\rm{inc}}\equiv S_{\rm{inc}}/2$ and $F^{+}_{\rm{inc}}=F^{-}_{\rm{inc}}\equiv F_{\rm{inc}}/2$, where $S_{\rm{inc}}$ and $F_{\rm{inc}}$ are the magnitudes of the total incident power and chiral flux, respectively. In fact, because the incident wave is linearly polarized, the total chirality flux is zero, i.e. $\mathbf{F}_{\rm{inc}} = \mathbf{F}^{+}_{\rm{inc}} + \mathbf{F}^{-}_{\rm{inc}} = 0$. However, because the individual fluxes have nonzero magnitude (and equal; they correspond to circularly polarized waves of equal amplitude), we define the incident flux magnitude as $F_{\rm inc} = |\mathbf{F}^{+}_{\rm{inc}}|+|\mathbf{F}^{-}_{\rm{inc}}|= 2|\mathbf{F}^{\pm}_{\rm{inc}}|\equiv 2F^{\pm}_{\rm{inc}}$.\\
\indent Next, we can associate the quantities related to the incident and reflected $+/-$ components as,
\begin{equation}
F^{\pm}_{\rm{inc}}=\mp \frac{\omega}{c}S^{\pm}_{\rm{inc}},\quad F^{\pm}_{\rm{refl}}=\mp \frac{\omega}{c}S^{\pm}_{\rm{refl}},
\end{equation}
and hence
\begin{equation}
\frac{F^{\pm}_{\rm{refl}}}{F^{\pm}_{\rm{inc}}}=\frac{S^{\pm}_{\rm{refl}}}{S^{\pm}_{\rm{inc}}}\Rightarrow
\frac{F^{\pm}_{\rm{refl}}}{F_{\rm{inc}}}=\frac{S^{\pm}_{\rm{refl}}}{S_{\rm{inc}}}\equiv R_{\pm},
\end{equation}
or simply $F_{\pm}/F_{\rm{inc}}=R_{\pm}$ as we use in the main manuscript (see Fig.\,9 main text). \\
\indent Furthermore, using $F_{\pm}/F_{\rm{inc}}=R_{\pm}$ it follows that:
\begin{equation}
\delta F= \frac{F_{+}-F_{-}}{F_{+}+F_{-}}= \frac{R_+ F_{\rm{inc}}-R_{-}  F_{\rm{inc}}}{R_+ F_{\rm{inc}}+R_{-}  F_{\rm{inc}}}= \frac{R_{+}-R_{-}}{R_{+}+R_{-}}=\rho_{\rm{DR}},
\end{equation}
i.e. the differential chirality flux $\delta F$ is equal to $\rho_{\rm{DR}}$, as we also demonstrate in Fig.\,9 of the main text.
\begin{figure}[t!]
\begin{center}
		\includegraphics[width=0.7\linewidth]{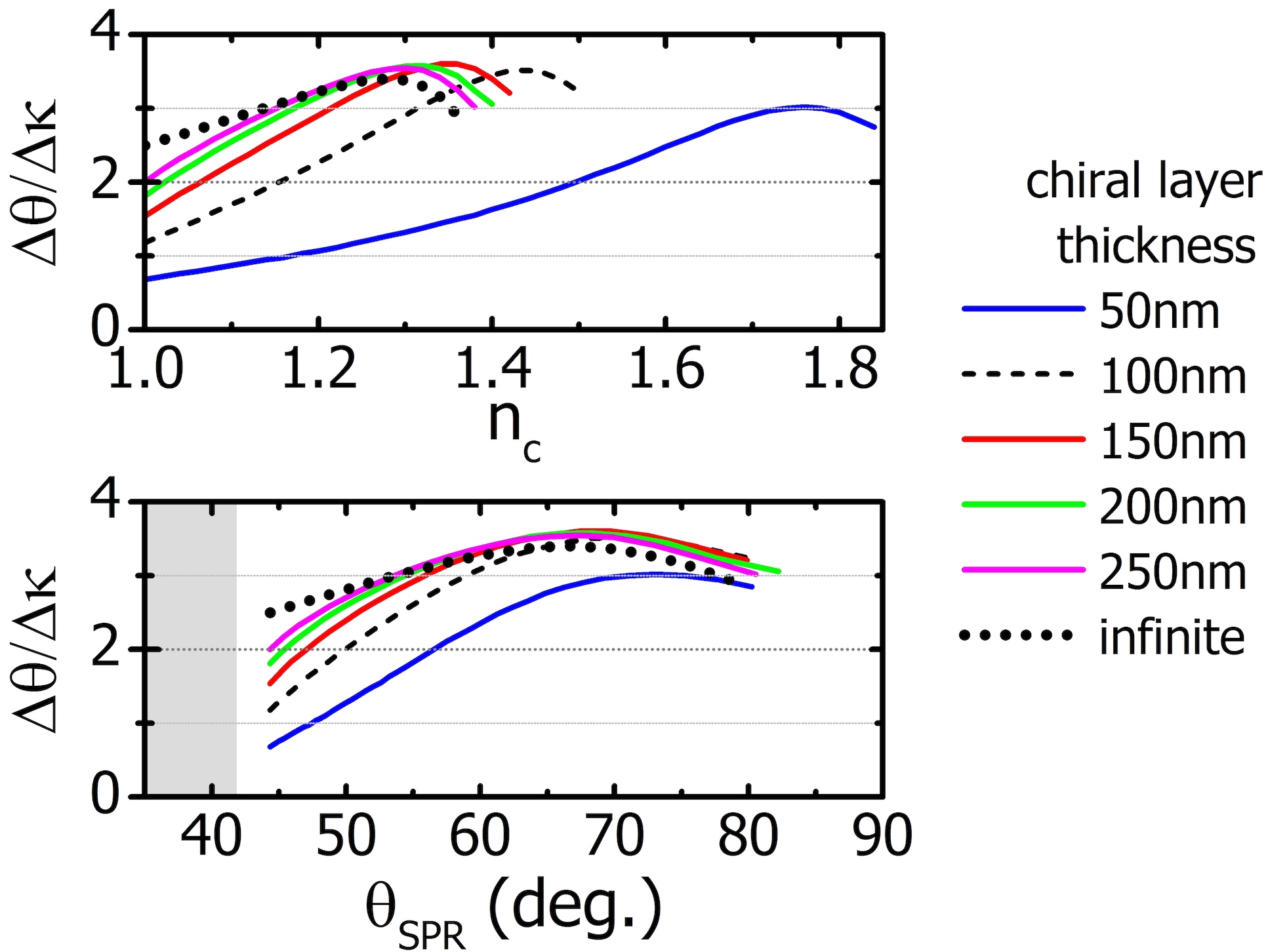}
	\caption{\small{$\Delta\theta/\Delta\kappa$ as a function of the chiral-layer thickness. The dashed line corresponds to the results shown in Fig.\,3b of the main text (reproduced here for easier comparison), and the dotted line denotes the limit for infinite chiral layer. The results are shown up to approximately 80\,deg. where the SPR dip is relatively pronounced.}}	
    	\label{fig:figA1}
\end{center}
\end{figure}
\begin{figure}[t!]
\begin{center}
		\includegraphics[width=0.8\linewidth]{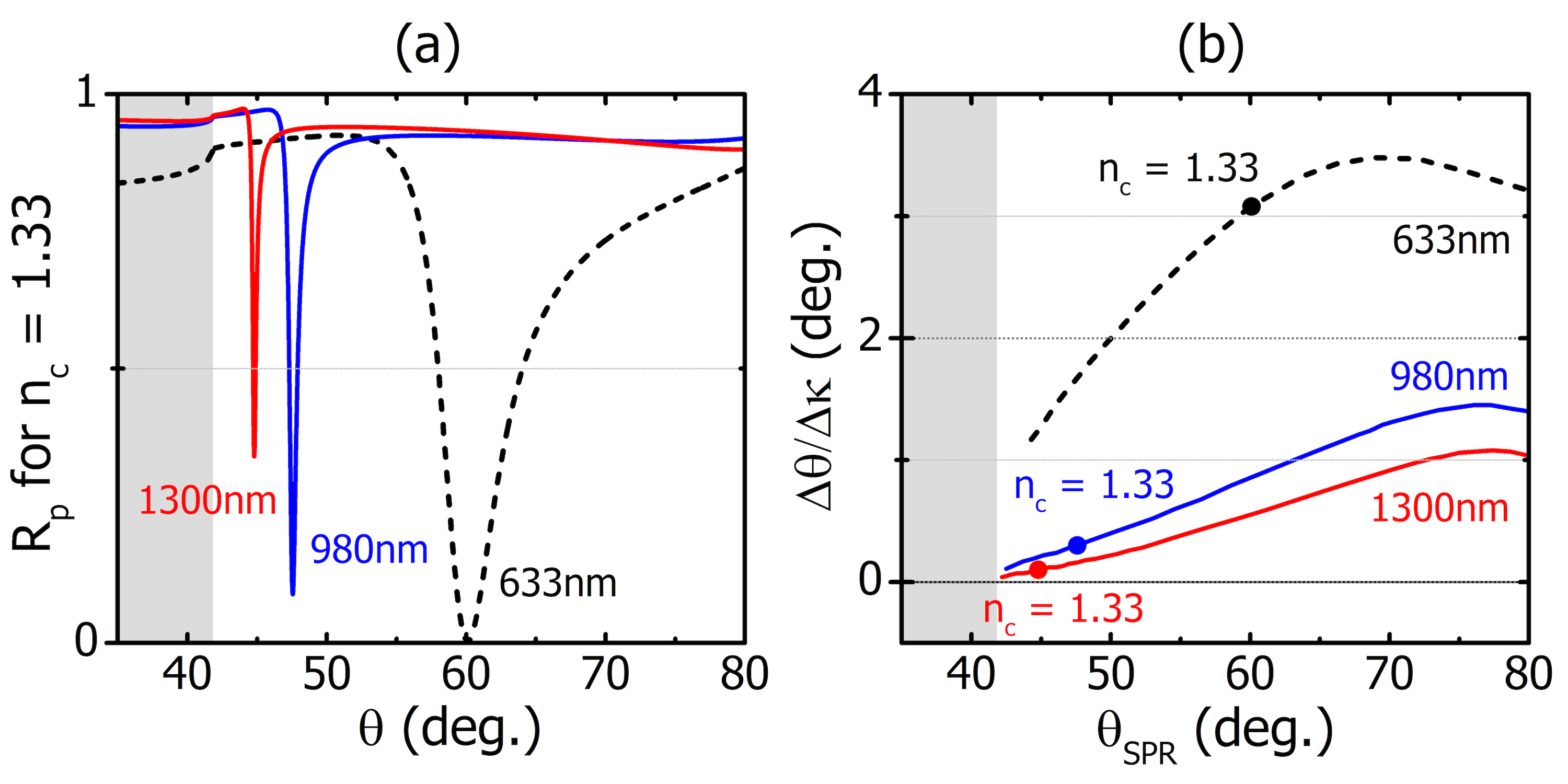}
	\caption{\small{Measurement sensitivity of $\Delta\theta$ of a 100\,nm chiral layer and 50\,nm Au layer for operation at 633\,nm (black dashed lines), 980\,nm (blue solid lines) and 1300\,nm (red solid lines). (a) $R_p$ reflectance for $n_c = 1.33$ and $\kappa = 0$. (b) $\Delta\theta/\Delta\kappa$ vs SPR angle. The cases for $n_c = 1.33$ shown in (a) are marked with dots of the same colour. The dashed black lines correspond to the results shown in Fig.\,3b of the main text.}}	
    	\label{fig:figA2}
	\end{center}
\end{figure}
\begin{figure}[t!]
\begin{center}
		\includegraphics[width=0.9\linewidth]{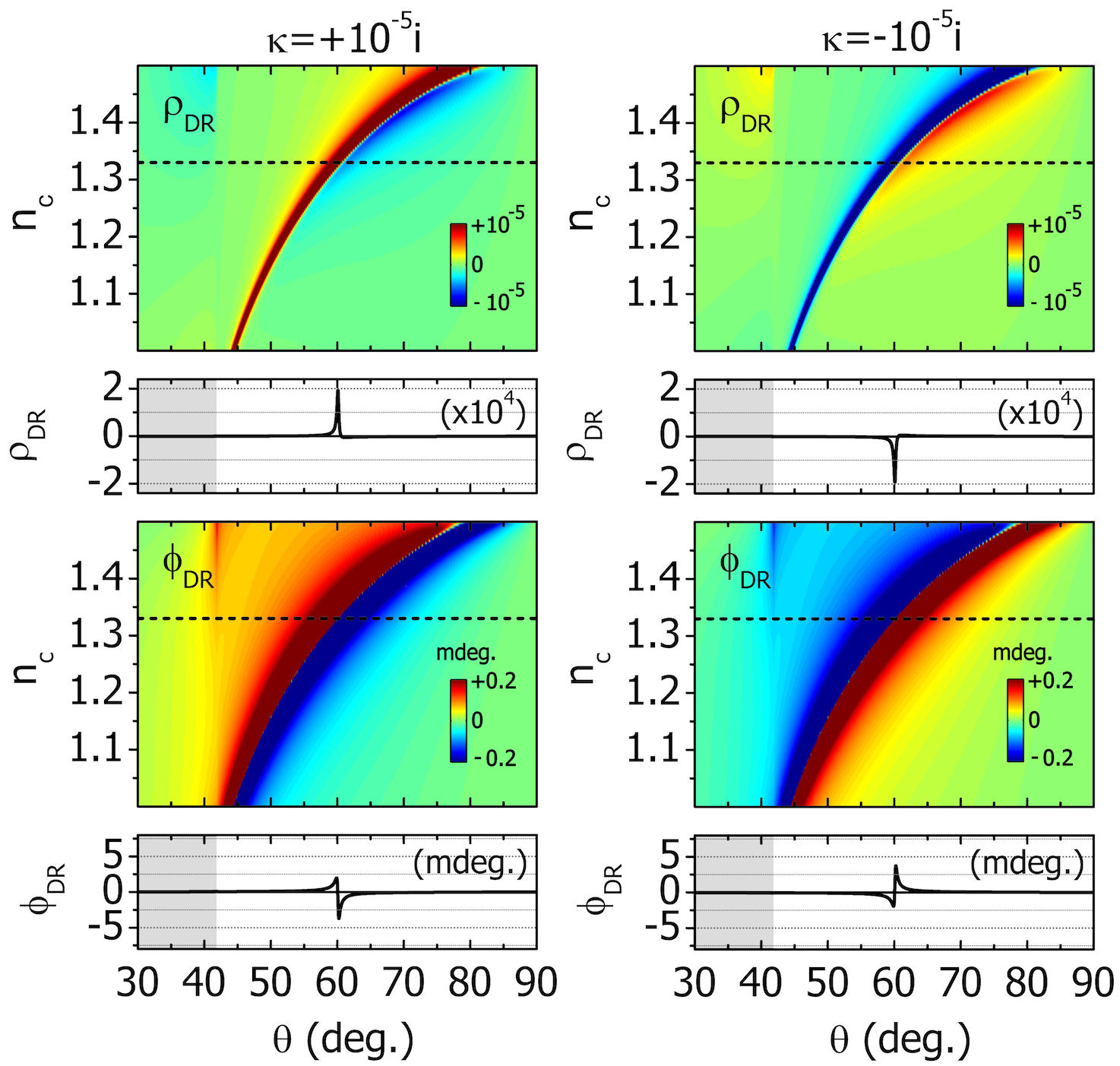}
	\caption{\small{Differential reflectance (DR) signals for a 100\,nm thin chiral layer of purely imaginary chiral parameter $\kappa = \pm10^{-5}i$, as function of $n_c$. To preserve the passivity we use Im($n_c$) $= 10^{-3}i$. Specific examples for $n_c = 1.33$, are marked with a horizontal dashed line and are shown separately below each panel.}}	
    	\label{fig:figA3}
	\end{center}
\end{figure}
\subsection{Sensitivity vs layer thickness}
In Fig.\,\ref{fig:figA1} we repeat the calculations of the measurement sensitivity ($\Delta\theta/\Delta\kappa$) presented in Fig.\,3b (main text), for chiral layers of variable thickness. We present the results in terms of $n_c$ and the corresponding SPR angle. We see that, with increasing chiral layer thickness, the measurement sensitivity converges to the limit of chiral substances of theoretically infinite extent (practically referring to electrically thick samples). In particular, for small SPR angles this increase is monotonic, but for large SPR angles the sensitivity reaches a maximum level for a thickness of $\sim$100\,nm-150\,nm, beyond which it gradually drops until convergence. Thus, due to the evanescent character of the SPP wave inside the chiral region, equivalent levels of sensitivity can be achieved for a large range of chiral-layer thicknesses. 
\subsection{SPR measurements vs. optical wavelength}
Our study, which has been focused on operation at 633\,nm, provides general conclusions which are applicable to any operation wavelength that supports pronounced SPRs. However, a change of the operation wavelength will induce changes in the metal permittivity and, thus, the SPR dispersion, which may significantly change the SPR reflectance. Therefore, in order to maintain the optimum coupling of the incident wave to the SPP wave, the thickness of the metal layer must be modified accordingly.\\
\indent To demonstrate the effect of the coupling strength on the measurement sensitivity, we examine the $\Delta\theta/\Delta\kappa$ dependence for three different SPR optical wavelengths (633\,nm, 980\,nm and 1300\,nm; typical optical wavelengths used in commercial SPR instruments\,\footnote{Schasfoort, R. B. M., Ed. Handbook of Surface Plasmon Resonance; The Royal Society of Chemistry, 2017.}), without changing the metal layer thickness, which is 50\,nm. As a result, the non-optimized metal thickness manifests as less pronounced SPR dips and reduced sensitivity in $\Delta\theta/\Delta\kappa$, accordingly. We show our results in Fig.\,\ref{fig:figA2}. 

\subsection{Differential measurements for Im($\kappa$)$\neq0$}
\begin{figure}[h]
\begin{center}
		\includegraphics[width=0.9\linewidth]{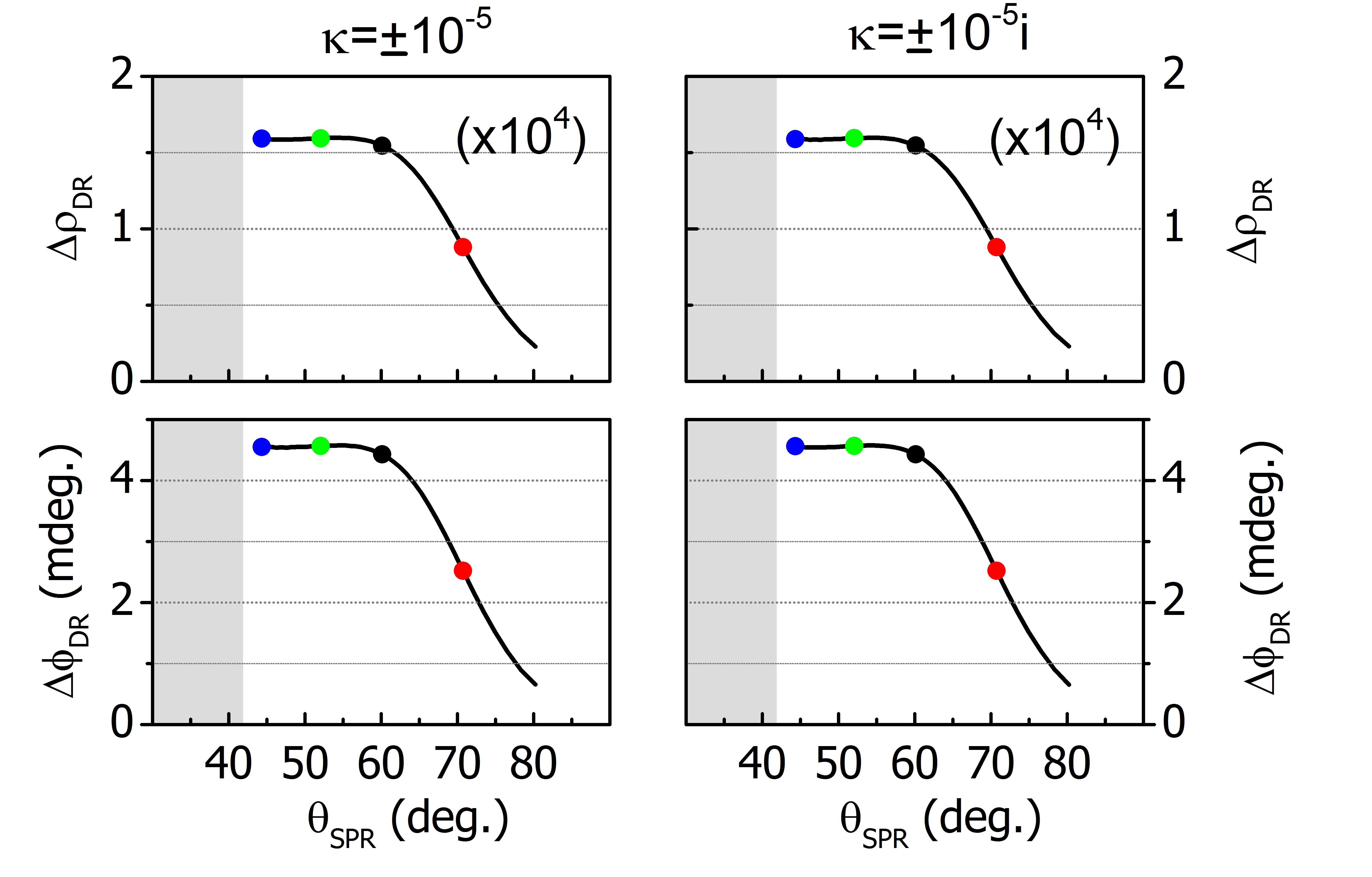}
	\caption{\small{Peak-to-peak values of the differential signals $\rho_{\rm{\small DR}}$ and $\phi_{\rm{\small DR}}$, $\Delta\rho_{\rm{\small DR}}$ (top), and $\Delta\phi_{\rm{\small DR}}$ (bottom) as function of the SPR angle [tuned via Re($n_c$)], for $\kappa = \pm10^{-5}$ (left) and $\kappa = \pm10^{-5}i$ (right). To preserve the passivity we use Im($n_c$) $= 10^{-3}i$.}}	
    	\label{fig:figA4}
\end{center}
\end{figure}

In Fig.\,8 of the main text we present differential reflectance signals for the case of purely imaginary $\kappa$, in particular for $\kappa =\pm 10^{-5}i$. These results are obtained for $n_c = 1.33 + 10^{-3}i$. For completeness, the calculations for variable Re($n_c$) and constant Im($n_c$)$= 10^{-3}i$ are shown in Fig.\,\ref{fig:figA3}. The peak-to-peak values of the differential reflectance signals, $\Delta\rho_{\rm{\small DR}}$ and $\Delta\phi_{\rm{\small DR}}$, for both purely real and pure imaginary $\kappa$ are shown in Fig.\,\ref{fig:figA4}. In both cases we consider Im($n_c$)$ = 10^{-3}i$. We see that the measured signals for either purely real or purely imaginary $\kappa$ are practically equal.

\end{document}